\documentclass{IEEEtran}

\makeatletter
\def\endthebibliography{
	\def\@noitemerr{\@latex@warning{Empty `thebibliography' environment}}
	\endlist
}
\makeatother

\usepackage{cite}

\usepackage[pdftex]{graphicx}
\usepackage[caption=false,font=footnotesize]{subfig}

\usepackage{amsmath}
\usepackage{amsfonts}
\usepackage{bm}
\usepackage{filecontents}
\usepackage{amssymb}
\usepackage{color}

\usepackage{epsfig}
\usepackage{verbatim}
\usepackage{url}

\usepackage{algorithm, tabularx}

\usepackage{lettrine}

\usepackage{lipsum}

\usepackage{siunitx}

\usepackage{soul,color}

\usepackage{mathrsfs}

\usepackage{array}

\usepackage[inline]{enumitem}

\usepackage{flushend}

\usepackage{xcolor}

\usepackage{dblfloatfix} 

\usepackage{algorithm}

\usepackage[noend]{algpseudocode}

\usepackage{setspace}
\usepackage{multirow}
\usepackage{array}
\newcolumntype{L}[1]{>{\raggedright\let\newline\\\arraybackslash\hspace{0pt}}m{#1}}
\newcolumntype{C}[1]{>{\centering\let\newline\\\arraybackslash\hspace{0pt}}m{#1}}
\newcolumntype{R}[1]{>{\raggedleft\let\newline\\\arraybackslash\hspace{0pt}}m{#1}}

\begin{document}
	
	\title{\LARGE ISI-Mitigating Character Encoding\\ for Molecular Communications via Diffusion}
	
	\author{Haewoong Hyun,~\IEEEmembership{Graduate Student Member,~IEEE}, Changmin Lee,~\IEEEmembership{Member,~IEEE},\\ Miaowen Wen,~\IEEEmembership{Senior Member,~IEEE}, Sang-Hyo Kim,~\IEEEmembership{Member,~IEEE}, and Chan-Byoung Chae,~\IEEEmembership{Fellow,~IEEE}
\thanks{The work is supported in part by the National Research Foundation of Korea (NRF) Grant through the Ministry of Science and ICT (MSIT), Korea (RS-2023-00208922, 2022R1A5A1027646).}
\thanks{The work of S.-H. Kim is supported in part by the Basic Science Research Program through the NRF of Korea funded by the MSIT under Grant NRF-2021R1A2C1008913.}
	\thanks{H.-W. Hyun, C.-M. Lee and C.-B. Chae are with the School of Integrated Technology, Yonsei University, Seoul 03722, Korea (e-mail: \{godnd980627, cm.lee, cbchae\}@yonsei.ac.kr).}
	\thanks{S.-H. Kim is with the Department of Electrical and Computer Engineering, Sungkyunkwan University, Suwon, Gyeonggi 16419, South Korea (email:  iamshkim@skku.edu).}
	\thanks{M. Wen is with the School of Electronic and Information Engineering, South China University of Technology, Guangzhou 510640, China (email: eemwwen@scut.edu.cn).}}
	
\markboth{IEEE Wireless Communications Letters, ~Vol.~XX, No.~XX, XXX~2023}{}
	
	\maketitle
	
	\begin{abstract}
		
%This letter presents a new algorithm for generating codebooks in molecular communications. The proposed algorithm is based on character entropy and aims to reduce inter-symbol interference (ISI) in molecular communication through diffusion. The proposed algorithm is based on Huffman coding and prevents, in the codebook, the occurrence of consecutive bit-1s. The error-correction process in the receiver removes the ISI in the slot right after bit-1. Further, we demonstrate through ISI analysis that the proposed algorithm decreases decoding error. The proposed codebook demonstrates improved performance in terms of character error rate compared to existing codebooks, as demonstrated through numerical analysis. Moreover, we verify the performance of the proposed algorithm on a real-time testbed.

This letter introduces a novel algorithm for generating codebooks in molecular communications (MC). The proposed algorithm utilizes character entropy to effectively mitigate inter-symbol interference (ISI) during MC via diffusion. Based on Huffman coding, the algorithm ensures that consecutive bit-1s are avoided in the resulting codebook. Additionally, the error-correction process at the receiver effectively eliminates ISI in the time slot immediately following a bit-1. We conduct an ISI analysis, which confirms that the proposed algorithm significantly reduces decoding errors. Through numerical analysis, we demonstrate that the proposed codebook exhibits superior performance in terms of character error rate compared to existing codebooks. Furthermore, we validate the performance of the algorithm through experimentation on a real-time testbed.

	\end{abstract}
	
	\begin{IEEEkeywords}
		Molecular communications, Huffman coding, inter-symbol interference, and testbed.
	\end{IEEEkeywords}
	
	\section{Introduction}

Molecular communication (MC) is a new communication paradigm that utilizes molecules as information carriers. This transfer of information can take place through various types of transmission, such as diffusion in a free space, transfer through gap junction channels, conveyance on a molecular rail using a motor like a microtubule, or propulsion by bacterial motors~\cite{tutorial}. In addition, there are various methods available to observers for decoding information once they receive messenger molecules. The most commonly used method is based on decoding the concentration of molecules~\cite{concentration}.

%When radio wave signals are significantly degraded or obstructed, relying on electromagnetic (EM) waves for communication becomes impractical due to the substantial propagation path loss. This is particularly true in underground or confined spaces \cite{EM,IoBNT}. MC is an alternative to traditional radio communication that utilizes molecules for transmission. MC has several advantages, including low power consumption, nano-level applicability, and biocompatibility, making it particularly useful in environments with severe EM degradation. With the growing attention on the healthcare industry, there is a need to deploy MC within the human body environment, and there is a growing demand for it \cite{Detection}.

In environments with degraded or obstructed radio wave signals, electromagnetic (EM) wave-based communication becomes impractical due to substantial path loss, especially in underground or confined spaces \cite{EM,IoBNT}. MC offers an alternative approach, utilizing molecules for transmission. MC brings several advantages, including low power consumption, nano-level applicability, and biocompatibility, making it highly useful in EM-degraded environments. As the healthcare industry gains attention, the demand for deploying MC within the human body environment is growing \cite{Detection}.

Previous research on MC has classified it into diffusion-based and non-diffusion-based categories. The most basic form of diffusion-based MC occurs in three-dimensional (3D) diffusion space, owing to its stochastic simplicity. However, this method suffers from severe inter-symbol interference (ISI) problems due to its long-tail channel-impulse response in a diffusive environment \cite{ISI}. As a result, it is essential to propose a new codebook that is suitable for MC and can overcome the inadequacy of existing codebooks \cite{ITA-2}. To this end, this letter presents a codebook-generating algorithm that minimizes ISI errors.

\begin{figure}[t]
	\begin{center}
		\includegraphics[width=0.90 \columnwidth,keepaspectratio]
		{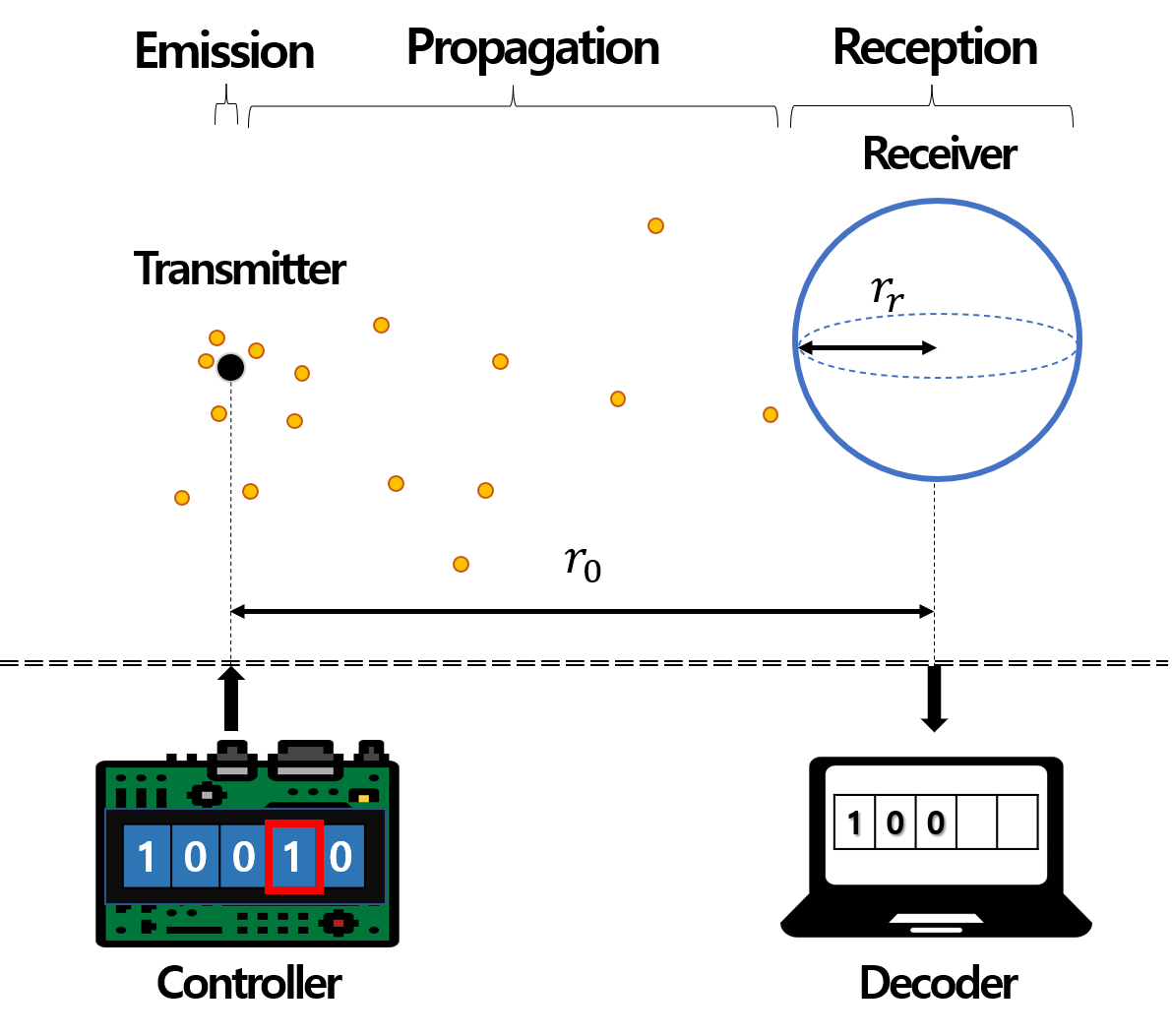}
		\vspace {-10pt}
		\caption{System model of molecular communication between a point transmitter and a fully absorbing spherical receiver.}
		\label{fig_system_model}
	\end{center}
	\vspace {-20pt}
\end{figure}

%The coding technique for MC comprises channel coding and source coding. In the field of channel coding techniques for MC, researchers have conducted studies that consider the characteristics of MC channels with severe ISI\cite{channel-coding1, channel-coding2}. For source-coding techniques of MC, the authors of \cite{Source-coding} studied a modified inverse source-coding scheme. The scheme, tailored to diffusion-based MC systems, was designed to control the percentages of bit-1 and bit-0 in the transmitted data. Compared to their work with channel coding in MC, though, researchers still need to study the area of source-coding. Hence, we focus on source-coding for MC and present a source-coding algorithm for MC via diffusion.

The coding technique for MC consists of source-coding and channel coding. In the area of channel-coding techniques for MC, researchers have conducted studies that consider the characteristics of MC channels with severe ISI \cite{channel-coding1, channel-coding2}. Regarding source-coding techniques for MC, the authors of \cite{Source-coding} proposed a modified inverse source-coding scheme that is specifically designed for diffusion-based MC systems. This scheme aims to control the percentage of bit-1 and bit-0 in the transmitted data. However, researchers need to further investigate the area of source-coding in MC, as it has not been explored to the same extent as channel coding. Thus, this letter focuses on source-coding for MC and presents a source-coding algorithm for MC via diffusion.

%A lossless data-compression scheme can remove statistical redundancy, a scheme known as entropy coding. When engineers consider the entropy of characters as they design a codebook, they minimize the expectation of the number of codebook’s bits. Consequently, this also affects data rates\cite{Thomas}. When it comes to data rates and power consumption, a codebook that employs entropy coding exhibits superior performance. Therefore, we propose a codebook-generating algorithm based on Huffman coding, which considers the frequency of characters but also the advantage of the decoding process, such as unique decodability\cite{Huffman}.

A lossless data-compression scheme known as entropy coding can remove statistical redundancy. Engineers designing a codebook consider the entropy of characters to minimize the expected number of bits in the codebook, which also affects data rates \cite{Thomas}. A codebook that employs entropy coding exhibits superior performance in terms of data rates and power consumption. Therefore, we propose a codebook-generating algorithm based on Huffman coding, which takes into account the frequency of characters as well as the advantages of the decoding process, such as unique decodability \cite{Huffman}.

ITA-2 (International Telegraph Alphabet 2) is a commonly employed teleprinter code that utilizes 5-bit codewords to represent the alphabet. However, in the context of MC, the presence of consecutive bit-1s in ITA-2 can lead to significant ISI. Moreover, ITA-2 lacks efficiency from an information theory perspective as it does not consider character frequency. On the other hand, Huffman coding, a variable-length code used for lossless data compression~\cite{Huffman}, takes character frequency into account. However, the occurrence of adjacent bit-1s in Huffman codewords renders it unsuitable for MC applications. Hence, we propose a codebook that eliminates consecutive bit-1s. This codebook is generated for data compression with character frequency taken into account and offers reduced ISI in MC scenarios.

%The main contribution of this letter is a novel scheme that generates a codebook with entropy coding and employs error-correction techniques to mitigate ISI effects in molecular communication via diffusion by exploiting the channel characteristics of MC. This is significant in that it achieved ISI mitigation in molecular communication for the first time from a source-coding perspective. We conduct a performance analysis of the proposed codebook, comparing it against ITA-2, a commonly used practical codebook, and the Huffman codebook, which is the optimal binary codebook. We set the Huffman codebook as the upper bound for data rates, and our results show that the error rate of the proposed codebook is lower than that of the other codebooks at the same data rate. Furthermore, we conduct simulations and testbed experiments to demonstrate the performance of the proposed algorithm.
The primary contribution of this letter is the introduction of a novel scheme that utilizes entropy coding to generate a codebook and incorporates error-correction techniques to mitigate ISI effects in molecular communication via diffusion. This approach is significant as it achieves ISI mitigation in molecular communication from a source-coding perspective, marking a notable advancement in the field. We conduct a comprehensive performance analysis of the proposed codebook, comparing it to the widely-used ITA-2 codebook and the Huffman codebook, which represents the optimal binary codebook. Setting the Huffman codebook as the upper bound for data rates, our results demonstrate that the proposed codebook achieves lower error rates than the other codebooks at equivalent data rates. Additionally, we validate the performance of the proposed algorithm through simulations and testbed experiments.

%The remainder of this letter is organized as follows. Section~\RomanNumeralCaps{2} describes the system model structure, how molecules are used for communication, and how we formulate the problem. In Section~\RomanNumeralCaps{3}, we present the codebook-generating algorithm for ISI mitigation and the error-correction method on the receiver. Section~\RomanNumeralCaps{4} analyzes the results and presents the algorithm's application. Finally, in Section~\RomanNumeralCaps{5}, we conclude and suggest directions for future work.

	\section{System Model and Problem Formulation}
	
%	Considered in this letter are diffusion-based MC systems in 3D space. As depicted in Fig. \ref{fig_system_model}, we assume a point source transmitter (Tx) and a fully absorbing spherical receiver (Rx) communicate through diffusion. The messenger molecules move according to Brownian motion \cite{nakano2013molecular}. Each molecule operates independently from the others, and their positions are determined by time-dependent random variables.

%Molecules emitted from the point source propagate in the 3D environment via diffusion. The diffusion process is characterized by the diffusion coefficient $D$. When molecules hit the surface of the spherical receiver, it absorbs them and counts each received molecule only once.

%This letter utilizes the probability formula, as outlined in \cite{3D}, for calculating the probability of molecules reaching the receiver up to a specific time point, $t$. The formula is as follows:
This letter considers diffusion-based MC systems in a 3D space, as shown in Fig.~\ref{fig_system_model}, where a point source transmitter (Tx) and a fully absorbing spherical receiver (Rx) communicate via diffusion. The movement of messenger molecules follows Brownian motion \cite{nakano2013molecular}, and each molecule operates independently of the others, with their positions determined by time-dependent random variables.

As molecules are emitted from the point source, they propagate through the 3D environment via diffusion, characterized by the diffusion coefficient $D$. The spherical receiver absorbs molecules that hit its surface and counts each received molecule only once.

To calculate the probability of molecules reaching the receiver up to a specific time point $t$, this letter uses the probability formula outlined in \cite{3D}, given as follows:
\begin{align}
\begin{split}
F_\text{hit} (t)=  \frac{r_r}{r_0} \, \text{erfc} \left[\frac{r_0-r_r}{\sqrt{4Dt}}\right]
\end{split}
\label{eq_F_hit}
\end{align}
where $r_0$ represents the distance between the transmitter and the center of the receiver, and $r_r$ represents the radius of the receiver.

%This system utilizes an evenly divided symbol slot to count absorbed molecules. A pulse peak time ($t_peak$) is when channel impulse response is maximized after molecules are emitted. The receiver follows a binary on-off keying system, which interprets the received signal as bit-1 or bit-0 based on whether the number of molecules received during the symbol slot ($t_s$) exceeds a predetermined threshold value, respectively.
This system employs a symbol slot that is evenly divided to count absorbed molecules, with a pulse peak time ($t_\text{peak}$) occurring when the channel impulse response is maximized after molecules are emitted. The receiver utilizes a binary on-off keying system that interprets the received signal as either bit-1 or bit-0 based on whether the number of molecules received during the symbol slot ($t_s$) exceeds a predetermined threshold value.

%The channel coefficients, which denote the probability of absorption for a molecule at the symbol slot after releasing, are given as $a_{k} =  F_\text{hit}(kt_s) - F_\text{hit}((k-1)t_s), \; k = 1, 2, ..., M, $ where $M$ is the channel memory. In the MC situation, we set the symbol duration for descending-order channel coefficients, i.e., $a_{1} > a_{2} >...> a_{M}$. It ensures that the first time slot receives more molecules than the others. If we set $t_s$ as a rule above, $a_4$ becomes smaller than 0.03, and $a_1 + a_2 + a_3$ is greater than 0.21. Because $a_4$ is a sufficient value to ignore if we set $t_s$ longer than the minimum duration that satisfies the rule, we set the channel memory to 3. Setting the channel memory to 3 is appropriate since it allows us to ignore $a_4$ if we set $t_s$ longer than the minimum duration required by the rule.

The channel coefficients, which indicate the probability of molecule absorption at the symbol slot after release, are given by $a_{k} = F_\text{hit}(kt_s) - F_\text{hit}((k-1)t_s)$, where $k = 1, 2, ..., M$, and $M$ is the channel memory. In the MC channel, we set the symbol duration for the channel coefficients of decreasing order, i.e., $a_{1} > a_{2} >...> a_{M}$, ensuring that the first time slot receives more molecules than the others. If we set $t_s$ accordingly, $a_{11}$ becomes smaller than 0.008, and $F_\text{hit}(10t_s)$ is greater than 0.33. We can safely ignore the effect of $a_i$ for $i>10$ if we set $t_s$ longer than the minimum duration required by the rule, and therefore, we set the channel memory to 10. It can be applied regardless of distance between transceiver, receiver size and diffusion coefficient in the environment assumed in this paper because the shape of the channel impulse response is maintained even if several parameters change.

	\section{Proposed codebook-generating algorithm}
	\label{pro}
	
Messenger molecules are utilized by MC for communication, and the arrival of molecules after the synchronized symbol slot can have an impact on the subsequent period, resulting in a significant challenge in reducing the effects of ISI. Furthermore, codebooks with shorter expected codeword lengths tend to exhibit better performance in terms of data rate. To tackle these issues, we propose a codebook that takes into account both the entropy of the transmitting characters and the mitigation of ISI effects. The proposed algorithm consists of two steps: source-coding and error-correction.

%	MC uses messenger molecules for communication. The molecules that arrive after the synchronized symbol slot affect the subsequent period. Reducing the effects of ISI is a significant challenge. Further, the codebook, which has a short expectation of codeword length, shows better performance in terms of data rate. To address these issues, we propose a codebook that mitigates ISI effects and regards the entropy of the transmitting characters. The proposed algorithm consists of two steps: source-coding and error-correction.

\subsection{Proposed codebook-generating algorithm}

\begin{algorithm}
	\caption{The proposed codebook-generating algorithm}
	\label{The codebook generating algorithm}
	\begin{algorithmic}[1]
		\Require{The set of prob. of appearance of characters: $C$}
			\State $n =\left\vert C \right\vert$
			\State $Q = C$ \Comment{min-priority queue $Q$}
			\For i = 1 to n-1
				\State $allocate \ a \  new \ node \ z$
				\State $z.left = x = \mathbf{EXTRACT-MIN(}$$Q$$\mathbf{)}$
					\State \Comment{Give bit-10 to the branch}
				\State $z.right = y = \mathbf{EXTRACT-MIN(}$$Q$$\mathbf{)}$
					\State \Comment{Give bit-0 to the branch}
				\State $z.prob = x.prob + y.prob$
				\State $\mathbf{INSERT(Q,z)}$
			\EndFor
			\State \textbf{return} $\mathbf{EXTRACT-MIN(}$$Q$$\mathbf{)}$
				\State \Comment{return the root of the tree}
	\end{algorithmic}
\end{algorithm}

We develop an algorithm that ensures that adjacent bit-1s do not occur in the codebook. To achieve this, we modify the Huffman coding algorithm by assigning bit-0 to the higher probability branch, which helps to reduce the average number of bit-1s in the coded sequence. This reduces the power consumption in MC and lowers the probability of errors. Our proposed codebook replaces bit-1 in the Huffman code with bit-10 to prevent adjacent bit-1s. This algorithm can also prevent consecutive bit-1s between characters. The set of characters $C$ is defined, where each character $c\in C$ is an object with an attribute $c.prob$ representing the appearance probability of the character $c$.

%This algorithm prevents the occurrence of adjacent bit-1s in the codebook. To minimize the codeword length, we take Huffman coding as the base algorithm and modify it to avoid consecutive bit-1s in a codeword. We assign bit-0 to the higher probability branch to reduce the average number of bit-1s in the coded sequence. It reduces power consumption in MC and provides the benefit of reducing the probability of error. The proposed codebook replaces bit-1 in the Huffman code with bit-10 to prevent an adjacent bit-1. This algorithm can also prevent the consecutive bit-1s between characters. $C$ is a set of characters, and each character $c \in C$ is an object with an attribute $c.prob$ which represents the appearance probability of the character $c$.

The codewords are determined by sequencing the bits in the paths from the root to leaf nodes, respectively. An example of a proposed codebook for the alphabet with the probability of characters in English, which follows \cite{Oxford-Dictionary}, is given in Section~\ref{ER}.
\begin{table}[t]
\caption{Proposed codebook for the alphabet}
\begin{center}
\label{codebooks}
{\footnotesize
\begin{tabular}{ccccc}
Alphabet & Huffman code & ITA-2 & Proposed code & Prob (\%) \\\hline
\hline
E & 101 & 10000 & 10010 & 11.2 \\
\hline
A & 0000 & 11000 & 0000 & 8.5 \\
\hline
R & 0001 & 01010 & 00010 & 7.6 \\
\hline
I & 0010 & 01100 & 00100 & 7.5 \\
\hline
O & 0100 & 00011 & 01000 & 7.2 \\
\hline
T & 0101 & 00001 & 010010 & 7.0 \\
\hline
N & 0110 & 00110 & 010100 & 6.7 \\
\hline
S & 1001 & 10100 & 100010 & 5.7 \\
\hline
L & 1100 & 01001 & 101000 & 5.5 \\
\hline
C & 1110 & 01110 & 1010100 & 4.5 \\
\hline
U & 00110 & 11100 & 0010100 & 3.6 \\
\hline
D & 01110 & 10010 & 01010100 & 3.4 \\
\hline
P & 01111 & 01101 & 010101010 & 3.2 \\
\hline
M & 10000 & 00111 & 100000 & 3.0 \\
\hline
H & 10001 & 00101 & 100010 & 3.0 \\
\hline
G & 11010 & 01011 & 10100100 & 2.5 \\
\hline
B & 11110 & 10011 & 101010100 & 2.1 \\
\hline
F & 001110 & 10110 & 001010100 & 1.8 \\
\hline
Y & 001111 & 10101 & 0010101010 & 1.8 \\
\hline
W & 110110 & 11001 & 1010010100 & 1.3 \\
\hline
K & 110111 & 11110 & 10100101010 & 1.1 \\
\hline
V & 111110 & 01111 & 10101010100 & 1.0 \\
\hline
X & 11111100 & 10111 & 10101010101000 & 0.3 \\
\hline
Z & 11111101 & 10001 & 101010101010010 & 0.3 \\
\hline
J & 11111110 & 11010 & 101010101010100 & 0.1 \\
\hline
Q & 11111111 & 11101 & 1010101010101010 & 0.1 \\
\hline\hline
Length & 4.2738 & 5 & 6.2491 \\
\hline
\end{tabular}
}
\end{center}
\vspace{-10pt}
\end{table}

	\subsection{Error Correction at Receiver}
	\label{ER}
\begin{algorithm}
	\caption{The error-correction process at the receiver}
	\label{The error correction process at receiver}
	\begin{algorithmic}[1]
		\State \text{Received signal sequence: r[$k$]} \Comment{$k$-bits length signal}
		\For i = 1 to $k$
			\If {i is not $k$ and r[i] is bit-1}
			\State $\text{Input bit-0 into r[i+1]}$
			\EndIf
		\EndFor
	\end{algorithmic}
\end{algorithm}

In this step, the error-correction algorithm is used to eliminate the ISI response to $a_2$. Although the proposed codebook ensures that no adjacent bit-1s occur in the transmitted codewords, which can mitigate ISI, it is important to note that ISI can still lead to decoding errors at the receiver, such as adjoining bit-1s in the decoded codewords. Therefore, the error-correction process corrects any such errors by forcing the bit right next to a decoded bit-1 to be a bit-0. This process is sequentially performed starting from the front of the received signals.

For example, if the received signal is 001\textbf{1}01\textbf{1}10, the error-correction process corrects it to 001\textbf{0}01\textbf{0}10 by altering the next bit after bit-1, in the third and the sixth place, to bit-0. This algorithm effectively mitigates the most significant ISI effect caused by the second time slot after releasing the molecules, which can improve the character error rate (CER) performance. Table \ref{codebooks} presents an example of our proposed codebook in comparison with the Huffman code.

%In this step, the error-correction algorithm eliminates the ISI response to $a_2$. The proposed codebook ensures that no adjacent bit-1s occur in the transmitted codewords, which can mitigate ISI. However, it is essential to note that ISI can still lead to decoding errors at the receiver, like adjoining bit-1s in the decoded codewords. Therefore, the error-correction process forces the right next to a decoded bit-1 with bit-0. This process is sequentially performed beginning at the front of the received signals.

%If the received signal is 001\textbf{1}01\textbf{1}10, the error-correction process corrects this signal to 001\textbf{0}01\textbf{0}10. This process alters the next bit after bit-1, in the third and the sixth place, to bit-0. This algorithm effectively mitigates the most significant ISI effect caused by the second time slot after releasing the molecules. Thus, it can improve the character error rate (CER) performance. An example of our proposed codebook in comparison with the Huffman code is presented in Table \ref{codebooks}.

\subsection{ISI Analysis}

	\begin{table}[t]
	\begin{center}
	\caption{Parameters used in the simulation}
	\renewcommand{\arraystretch}{1.1}
	\label{tbl_system_parameters}
	\begin{tabular}{p{5.5cm} l}
	\hline
	\bfseries{Parameter} 							& \bfseries{Value} \\ 
	\hline
	Diffusion coefficients ($D$) 	& $79.4\,\,\si{\micro\metre^2/\second}$\\
	Distances between Tx and Rx ($r_0$)			& $4\,\, \si{\micro\metre} $\\
	Receiver radius ($r_r$)			& $2\,\,\si{\micro\metre}$\\
	\hline
	\end{tabular} 
	\end{center}
	\renewcommand{\arraystretch}{1}
	\vspace{-18pt}
	\end{table}

%To establish the superiority of the proposed algorithms, we compare the expected ISI of the Huffman code and the proposed code. ISI can be determined by evaluating the combined interference caused by all preceding bit-1s on the current slot within the channel memory $M$. ISI terms for bit-1 help diminish the bit error rate; therefore, to analyze ISI effects on increasing error rate, we need to consider only the ISI terms for bit-0. ISI effects on bit-0 can be represented by a linear combination of the channel coefficients as 
To demonstrate the effectiveness of the proposed algorithms, we compare the expected ISI of the Huffman codebook with that of the proposed codebook. ISI, which is the interference caused by all preceding bit-1s on the current slot within the channel memory $M$, can be used to evaluate the effect of ISI on the bit error rate. To analyze the impact of ISI on the increasing error rate, we need to consider only the ISI terms for bit-0. The effects of ISI on bit-0 can be expressed as a linear combination of the channel coefficients, as follows:
\begin{align}
\begin{split}
E[\text{ISI}_{0}] = p_0\sum_{i=1}^{K} \sum_{j=2}^M a_j {p^i} {s^i}_{M-j+1}
\end{split}
\label{ISI}
\end{align}
where $p_0$ is the appearance probability of bit-0 in the code- book, $K$ is the number of all possible sequences transmitted before the bit-0 slot, ${s^i}_k$ is the $k$-th bit of the $i$-th sequence of all possible sequences with length $M$ containing the current bit-0, and $p^i$ is the probability of $i$-th sequences in all possible sequences\cite{ISI-mitigating}.

We define ${N_0}^H$ as the number of possible sequences that affect the current bit-0 slot. In (\ref{ISI_h}), ${p_h}^i$ is the probability of the $i$-th sequences of all possible sequences of the Huffman codebook. The expectation of ISI for bit-0 of the Huffman codebook for the current slot is as follows:
\begin{align}
\begin{split}
E[{\text{ISI}_{0}}^{H}] = p_0\sum_{i=1}^{{N_0}^{H}} \sum_{j=2}^M a_j {p_h}^i {s^i}_{M-j+1}.
\end{split}
\label{ISI_h}
\end{align}
Assume that the number of all possible sequences before bit-0 for the proposed codebook with channel memory $M$ is ${N_0}^P$. Therefore, the expectation of ISI of the proposed codebook for the current bit-0 slot is as follows:
\begin{align}
E[{\text{ISI}_{0}}^{P}] = p_0\sum_{i=1}^{{N_0}^P} \sum_{j=3}^M a_j {p_p}^{i} {s^i}_{M-j+1}
\label{ISI_p}
\end{align}
%where ${p_p}^{i}$ is the proportion of $i$-th sequences in all possible sequences before bit-0 for the proposed codebook. In the proposed codebook case, the error-correction removes ISI effect of adjacent bit-1s. Therefore, $E[{\text{ISI}_{0}}^{P}]$ has no $a_2$ term.
where ${p_p}^{i}$ represents the proportion of the $i$-th sequence among all possible sequences before the occurrence of bit-0 in the proposed codebook. In the case of the proposed codebook, error correction effectively eliminates the ISI caused by adjacent bit-1s. As a result, $E[{\text{ISI}_{0}}^{P}]$ does not include the $a_2$ term.

%%%%

In this letter, we calculated that $E[{\text{ISI}_{0}}^{H}]$ is equal to $0.2719a_2 + 0.2745a_3$ and $E[{{\text{ISI}}_{0}}^{P}]$ is equal to $0.1904a_3$, for the alphabet codebook based on \cite{Oxford-Dictionary} and $M=3$. This letter presents an algorithm to remove the effect of $a_2$, a channel coefficient that has the greatest impact on ISI, so we analyze ISI based on channel memory 3. These values were obtained using the channel coefficients given in Table~\ref{tbl_system_parameters} and equations (\ref{ISI}) and (\ref{ISI_h}). %Since $a_2$ is at least twice $a_3$ according to the setting of $t_s$, $E[{{\text{ISI}}_{0}}^{P}]$ is smaller than $E[{{\text{ISI}}_{0}}^{H}]$.
For any arbitrary alphabet and its probability profile, the proposed codebook generally prevents the effect of $a_2$ on $E[{\text{ISI}_{0}}^{P}]$. As a result, the coefficients of ISI terms, represented by $a_i$ in $E[{{\text{ISI}}_{0}}^{P}]$, are lower than the coefficients of both 0.5 for the uncoded case and the Huffman codebook. This implies that the proposed codebook has a lower expected ISI for a bit-0 compared to both the uncoded case and the Huffman codebook. This result is maintained in a larger channel memory environment.

%	In this letter, $E[{\text{ISI}_{0}}^{H}]$ is $0.3060a_2 + 0.4668a_3$ and $E[{{\text{ISI}}_{0}}^{P}]$ is $0.6364a_3$ for the alphabet codebook about \cite{Oxford-Dictionary} and $M = 3$. These terms are obtained from (\ref{ISI}) and (\ref{ISI_h}) with the channel coefficients, decided by the parameters given in Table~\ref{tbl_system_parameters}. According to the setting of $t_s$ above, $a_2$ is at least twice $a_3$. Thus in the above formula, $E[{{\text{ISI}}_{0}}^{P}]$ is smaller than $E[{{\text{ISI}}_{0}}^{H}]$. For an arbitrary alphabet and its probability profile, the proposed code prevents the effect of $a_2$ on $E[{\text{ISI}_{1}^{P}}]$ in general. Therefore, the coefficients of ISI terms, $a_i$’s in $E[{{\text{ISI}}_{0}}^{P}]$, are lower than the coefficients 0.5 of the uncoded case and the Huffman’s. This result implies that the proposed codebook has a lower expected ISI for a bit-0 compared to both the uncoded case and the Huffman codebook.

\subsection{Threshold Decision}
\label{Th}

We utilize the adaptive threshold proposed in \cite{ISI-mitigating} for conventional codebooks in this letter. However, as mentioned earlier, the proposed codebook is not affected by the bit-10 case, and thus, an appropriate threshold must be set for it. Let the number of molecules emitted per bit-1 be $N$, and let bit-$X$ denote either bit-0 or bit-1.
The CER of the proposed codebook ($\text{CER}_{P}$) is based on the probabilities of two scenarios: if the transmitted signal in the last three slots is bit-$X$01, the receiver determines the current bit to be bit-0, and if the transmitted signal is bit-100, the receiver decides the current bit to be bit-1. Let $\mathbb{X}=[X_{i-2},X_{i-1},X_i]$ represent the transmitted signals for the $(i-2)$-th, $(i-1)$-th, and $i$-th time slots, and let $Y_i$ represent the received signal for the $i$-th time slot. The CER of the proposed codebook can then be expressed as follows:
\begin{align}
\begin{split}
\text{CER}_{P} =  &\text{Pr}[Y_i=1 | \mathbb{X}=[1,0,0]] \\
			 &+\text{Pr}[Y_i=0 | \mathbb{X}=[X,0,1]].
\label{CER_h}
\end{split}
\end{align}

\begin{figure}[t]
	\begin{center}
		\includegraphics[width=1\linewidth]{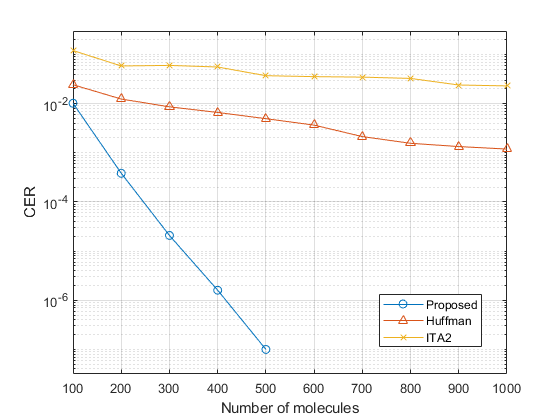}
		\end{center}
		\vspace{-10pt}
		\caption{CER versus the number of molecules per character for {ITA-2}, Huffman, and the proposed codebook. Here, we assume the alphabet appearance probability according to \cite{Oxford-Dictionary} at two characters per second.}
		\label{Alphabet_BER_CER}
		\vspace{-15pt}
\end{figure}
We assume that $R_i$ is the sequence of the number of molecules received in each time slot for the $i$-th codeword, except zero. And ${R_i}^{exc}$ is the sequence of slots in $R_i$ excluding the slots for bit-10. We obtain two values, $N_1=\min_{i}(\max(R_i))$ and $N_3=\mathbb{E}[\max({R_i}^{exc})]$, for threshold setting through pilot signals for all codewords. In this process, we balance the probabilities of the two error causes for the proposed codebook.
To minimize the error rate, the threshold should be the internally dividing point between $N_1$ and $N_3$ with a ratio of $\sqrt{N_1(1-N_1/N)}:\sqrt{N_3(1-N_3/N)}$. This approach is used to decide the threshold, as it approximately minimizes the error in the situation where molecules follow Brownian motion.
%, and the probability of incorrectly determining the current bit as bit-0 when the transmitted signal is bit-1 is lower at the lower threshold of at most $N{\cdot}a_1$. Conversely, when the receiver incorrectly determines the current bit as bit-1 when bit-0 is transmitted, the higher threshold of at least $N{\cdot}a_3$ has a lower error rate.

	\section{Simulation and Testbed Results}
	
	\subsection{Performance Evaluations}
	
	%We set the system parameters for simulations, as detailed in Table~\ref{tbl_system_parameters}. The performance of each codebook is evaluated based on the CER level, which is depicted in the graphs presented in this paper. We use the codebooks for the appearance probability of the alphabet which follows \cite{Oxford-Dictionary}.
	
	We established the system parameters for the simulations, as outlined in Table~\ref{tbl_system_parameters}, to evaluate the performance of each codebook based on the CER level. The graphs presented in this paper depict the results. The appearance probability of the alphabet follows \cite{Oxford-Dictionary} for the selection of codebooks.
	
	To ensure a fair comparison of the performance of each codebook in molecular communication, several components must be considered when transmitting the molecular signal. Firstly, to assess the performance of each codebook at the same data rate, we set the same character-transmitting duration, which represents the time required to send one character, and determine the symbol slot for each codebook based on its long-term average bit length. Secondly, we balance the number of emitting molecules per bit-1, which is considered the signal power for MC. The Huffman and proposed codebooks have the same long-term average number of bit-1s per character (1.9753 bit-1 per character), while the ITA-2 codebook has an average of 2.4696 bit-1s per character. To ensure a fair comparison among the different codebooks in terms of the number of emitted molecules per character, we employed a ratio of 1:1:0.7998 for bit-1 molecules in the simulation for the Huffman, proposed, and ITA-2 codebooks, respectively. By balancing these factors, we can ensure a fair and accurate comparison of the performance of each codebook in molecular communication.
	
	%Several components must be considered to transmit the molecular signal for a fair comparison. Firstly, we set the same character-transmitting duration, representing the time required to send one character. The long-term average bit length determines the symbol slot for each codebook. It makes the simulation assess the performance of each codebook at the same data rate. Secondly, we balance the number of emitting molecules per bit-1, which is considered the signal power for MC. The Huffman and proposed codebooks have the same long-term average number of bit-1s per character (1.9753 bit-1 per character), while the ITA-2 codebook has an average of 2.4696 bit-1s per character. To ensure a fair comparison among the different codebooks in terms of the number of emitted molecules per character, we employed a ratio of 1:1:0.7998 for bit-1 molecules in the simulation for the Huffman, proposed, and ITA-2 codebooks, respectively. By balancing these factors, we can ensure a fair and accurate comparison of the performance of each codebook in molecular communication.

	%We use the binomial distribution to determine the number of received molecules for each symbol slot in the simulation. Since we assume a fully absorbing receiver in the system model, the molecules counted in the previous slot are excluded from the subsequent slots. Thus, when the transmitter emits $N$ molecules, the number of received molecules at the receiver can be represented as follows: 
	We employ the binomial distribution to determine the number of molecules received for each symbol slot in the simulation. As we assume a fully absorbing receiver in the system model, molecules counted in the previous slot are excluded from subsequent slots. Thus, if the transmitter emits $N$ molecules, the number of received molecules at the receiver is represented as follows:
$$n_i \sim \mathcal{B}(N- \Sigma_{k=1}^{i-1} n_k,  a_i/(1-\Sigma_{k=1}^{i-1} a_k))$$
where $n_i$ is the number of received molecules in the $i$-th slot after the emitting time slot.  In this study, we use the values from $n_1$ to $n_{10}$ in a single release of molecules for channel memory 10. To obtain statistically stable results, the simulation is repeated 1,000,000 times for each 10-length random alphabet sequence, and the reported results are the average values.

%In this work, we use the values from $n_1$ to $n_3$ in a single release of molecules for channel memory 3. The simulation results are the average values of 100,000 repetitions for each 20-length random alphabet sequence.

\begin{figure}[!t]
	\begin{center}
		\includegraphics[width=1\linewidth]{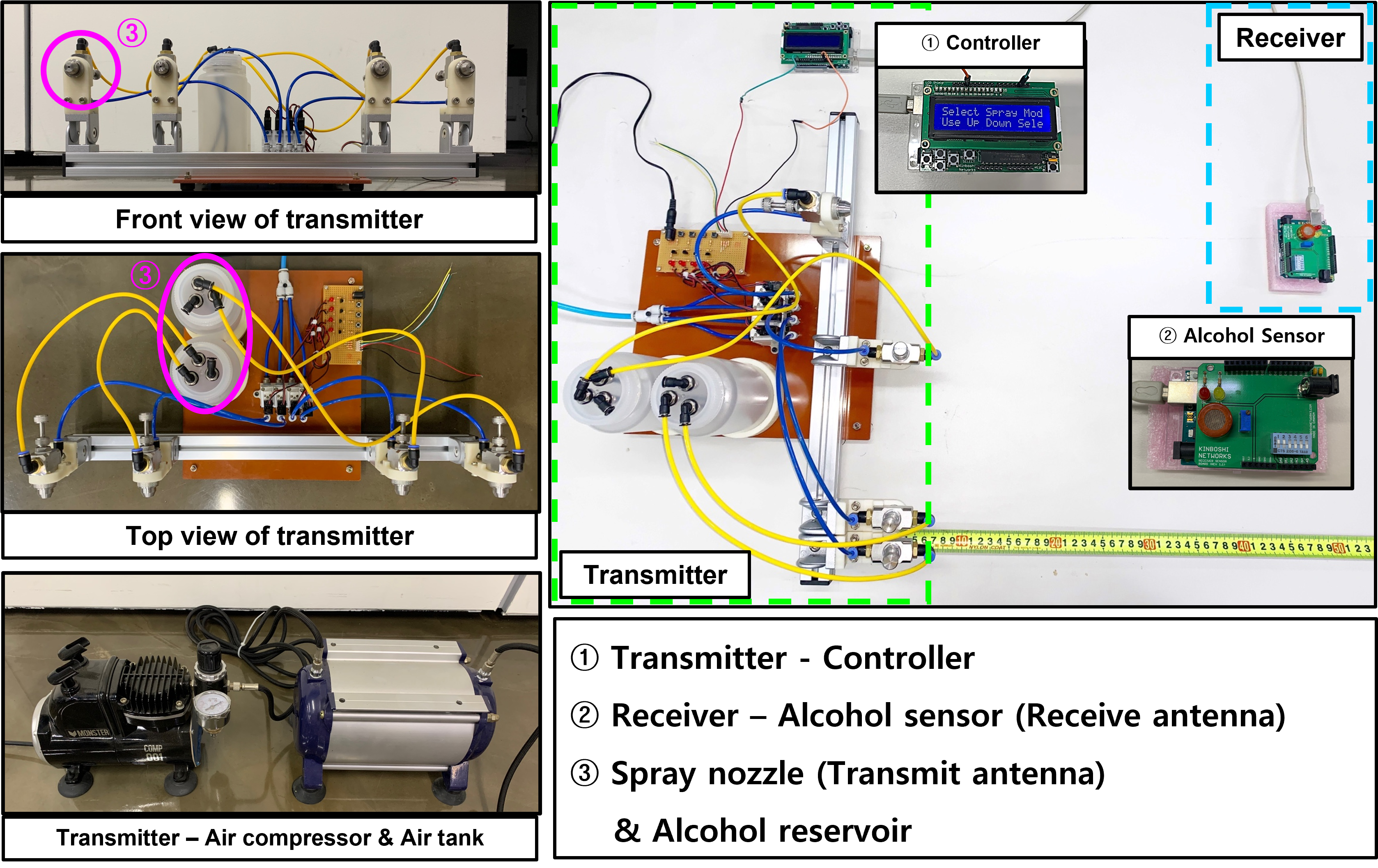}
		\end{center}
		\vspace{-10pt}
		\caption{Testbed configuration.}
		\label{Testbed}
		\vspace{-10pt}
	\end{figure}

Fig. 2 illustrates the relationship between the CER and the number of molecules per bit, with a transmission rate of two characters per second (cps), using the alphabet appearance probability from \cite{Oxford-Dictionary}. The conventional codebooks (Huffman and ITA-2) use the adaptive threshold \cite{ISI-mitigating}, while the proposed codebook uses the pilot signals as described in Section~\ref{Th} to determine the threshold. The x-axis represents the number of molecules per character in the simulation. The figure shows that the proposed codebook outperforms the conventional codebooks across the entire range of the number of molecules. Despite the longer symbol slot durations of the proposed codebook compared to the conventional codebooks, the proposed codebook achieves a lower CER due to the effect of error-correction at the receiver. The channel coefficient ratio of the substantive coefficient $a_1$ for transmitting a signal is much larger than that of other slots which belong to the ISI term, giving the conventional codebooks an advantage.

\begin{figure}[t]
	\begin{center}
		\includegraphics[width=1\linewidth]{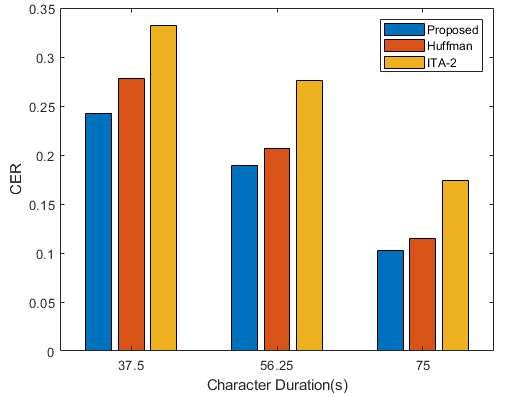}
		\end{center}
		\vspace{-15pt}
		\caption{CER of transmitting alphabet, which has an appearance probability according to \cite{Oxford-Dictionary}, using ITA-2, Huffman, and the proposed codebook in the testbed.}
		\label{Testbed}
		\vspace{-12pt}
\end{figure}

 	\subsection{Testbed Result}

We evaluated the performance of the proposed codebook through real-time experiments conducted on a macro-scale testbed, as illustrated in Fig. 3, comprising a transmitter and receiver. The air compressor condenses the air and delivers it to an air tank in the transmitter section. The controller manages the spray nozzle that is connected to the alcohol reservoir, which receives compressed air from the air tank. The nozzle sprays alcohol when the controller opens the spray nozzle. The receiver detects the emitted alcohol molecules using the alcohol sensor and decodes the received signal based on the selected codebook.

The distance between the transmitter and receiver was fixed at 1 meter. To ensure the descending order of the channel coefficients, we set the symbol slot duration to be longer than five seconds. Consequently, we evaluated the CER performance for character-transmitting durations of over 30 seconds. The number of molecules per character for each codebook was controlled by adjusting the spray duration, i.e., the duration in which the controller opens the spray nozzle to emit alcohol molecules. To ensure a fair comparison of power consumption among the three codebooks, i.e., the Huffman, the proposed, and the ITA-2, we set the spray duration for bit-1 at a 1:1:0.7998 ratio.

%As shown in Fig. 4, the proposed codebook yields a better CER performance than the existing codebooks across all character duration ranges. This result shows that the performance of the proposed codebook in the testbed is consistent with its simulation performance trend, albeit with a scaling difference in the channel impulse response due to the different scales of the two environments.

	The results presented in Fig. 4 indicate that the proposed codebook outperforms the existing codebooks in terms of CER across all character duration ranges, as confirmed through the real-time experiment conducted in the testbed. This result is consistent with the simulation performance trend of the proposed codebook, despite the scaling difference in the channel impulse response resulting from the different scales of the two environments. This proves the feasibility of the proposed algorithm in the long-tail channel situation.

 	\section{Conclusion}
In this letter, we proposed algorithms for generating a codebook and correcting errors that effectively mitigate ISI. Algorithm 1, which inserts bit-0, reduces error rates, but the symbol duration is shorter due to more bit-0s when compared to the Huffman and ITA-2 codebooks. The error-correction process removed incorrect bit errors immediately after bit-1, resulting in better CER performance than the existing codebooks. The proposed algorithm's performance was verified through simulation and on a real-time testbed, and it can be applied to other molecular communication (MC) situations that experience ISI due to a long-tail channel-impulse response, such as diffusion MC and drift-based MC with similar impulse response patterns. In future research, we intend to explore the impact of the appearance-frequency distribution of transmitting characters on the CER. We also plan to extend our work by considering arbitrary appearance probabilities of characters. Furthermore, we aim to compare the performance of our proposed algorithm with various other channel coding techniques for mitigating ISI.

\bibliographystyle{IEEEtran}
\bibliography{sample}

\end{document}